%% file: Main.tex
\documentclass{bmcart}

\usepackage[utf8]{inputenc} 

\usepackage{array}
\usepackage[cmex10]{amsmath}
\usepackage{amssymb}
\usepackage{graphicx}

\startlocaldefs
\endlocaldefs

\begin{document}

\begin{frontmatter}

\begin{fmbox}
\dochead{Research}


\title{A Line-of-Sight Channel Model for the 100--450 Gigahertz Frequency Band}


\author[
   addressref={aff1},                   
   corref={aff1},                       
   email={joonas.kokkoniemi@oulu.fi}   
]{\inits{JK}\fnm{Joonas} \snm{Kokkoniemi}}
\author[
   addressref={aff1},
   email={janne.lehtomaki@oulu.fi}
]{\inits{JL}\fnm{Janne} \snm{Lehtom\"{a}ki}}
\author[
   addressref={aff1},
   email={markku.juntti@oulu.fi}
]{\inits{MJ}\fnm{Markku} \snm{Juntti}}


\address[id=aff1]{
  \orgname{Centre for Wireless Communications (CWC), University of Oulu}, 
  \street{P.O. Box 4500},                     %
  \postcode{90014}                                
  \city{Oulu},                              
  \cny{Finland}                                    
}



\end{fmbox}


\begin{abstractbox}

\begin{abstract} 
\input{EUCNC19abstract.tex}

\end{abstract}




\end{abstractbox}
%

\end{frontmatter}




\section{Introduction}
    \label{section:1}
    \input{EUCNC19intro}

\section{Simplified Molecular Absorption Loss Model}
    \label{section:3}
    \input{EUCNC19simpl}

\section{Numerical Results and Discussion}
    \label{section:4}
    \input{EUCNC19results}

\section{Conclusion}
    \label{section:5}
    \input{EUCNC19concl}
    
\section{Methods/Experimental}

This paper is a purely theoretical model on simple way to estimate the absorption loss. Although theoretical, the original data obtained from the HITRAN database \cite{HITRAN12} is based on experimental data. The goal in this article is to simplify the complex database approach into simple polynomial equations with only few floating parameters, such as humidity and frequency. As such, the model produced in this paper is suitable for LOS channel loss estimation for various wireless communications systems. Those include back- and fronthaul connectivity and general LOS link channel estimation. The work is heavily based on the HITRAN database and the theoretical models for absorption loss as well as simple LOS free space path loss models.



\begin{backmatter}

\section*{Abbreviations}
    5G: fifth generation; 6G: sixth generation; B5G: beyond fifth generation; FSPL: free space path loss; HITRAN: high-resolution transmission molecular absorption database; ITU-R: International Telecommunication Union Radio Communication Sector; LOS: line-of-sight; mmWave: millimeter wave; Rx: receiver; Tx: transmitter.
    
\section*{Availability of data and materials}
    Not applicable.

\section*{Competing interests}
  The authors declare that they have no competing interests.

\section*{Funding}
    \input{EUCNC19ack.tex}

\section*{Author's contributions}
    JK derived the molecular absorption loss model. All the authors participated in writing the article and revising the manuscript. All the authors read and approved the final manuscript.

\section*{Authors' information}
Address of all the authors: Centre for Wireless Communications (CWC), University of Oulu, P.O. Box 4500, 90014 Oulu, Finland.

Emails: forename.surname(at)oulu.fi

\bibliographystyle{bmc-mathphys} 
\bibliography{References.bib}      

\end{backmatter}
\end{document}

%% file: EUCNC19abstract.tex
This paper documents a simple parametric polynomial line-of-sight channel model for 100--450 GHz band. The band comprises two popular beyond fifth generation (B5G) frequency bands, namely, the D band (110--170 GHz) and the low-THz band (around 275--325 GHz). The main focus herein is to derive a simple, compact, and accurate molecular absorption loss model for the 100--450 GHz band. The derived model relies on simple absorption line shape functions that are fitted to the actual response given by complex but exact database approach. The model is also reducible for particular sub-bands within the full range of 100--450 GHz, further simplifying the absorption loss estimate. The proposed model is shown to be very accurate by benchmarking it against the exact response and the similar models given by International Telecommunication Union Radio Communication Sector (ITU--R). The loss is shown to be within $\pm$2 dBs from the exact response for one kilometer link in highly humid environment. Therefore, the its accuracy is even much better in the case of usually considered shorter range future B5G wireless systems.

%% file: EUCNC19intro.tex
The high frequency communications aims at finding large contiguous bandwidths to serve high data rate applications and services. Especially the millimeter wave (mmWave) frequencies (30--300 GHz) are among the most prominent to provide high data rate connectivity in fifth generation (5G) and beyond (B5G) systems \cite{Rappaport2013,Rappaport2019,Latva-Aho2019}. In this context, the 5G systems will utilize the below 100 GHz frequencies, whereas the B5G systems, including the visioned sixth generation (6G) systems will look for spectral resources also above 100 GHz \cite{Latva-Aho2019}. These frequencies would theoretically allow very large bandwidths, but there are still many challenges to reach the above 100 GHz band efficiently with compact and portable devices.

Besides the hardware, knowledge of the operational channels are in the focal point to understand the fundamental physical limits of the transmission platform. This paper considers the line-of-sight (LOS) propagation in the sub-THz and low-THz frequencies at frequency range from 100 GHz to 450 GHz\footnote{This paper is an invited extended version of the conference paper presented in the EuCNC'19 conference \cite{Kokkoniemi2019}}. The main emphasis is on modelling the molecular absorption loss in this frequency interval. The molecular absorption loss is caused by the energy of the photons being absorbed by the free energy states of the molecules \cite{Jornet2011}. The absorption loss is described by the Beer-Lambert law and it causes exponential frequency selective loss on the signals as a function the frequency. The lowest absorption lines lie at low mmWave frequencies \cite{HITRAN12}, but the first major lines appear well above 100 GHz. 

The molecular absorption loss is most often modelled by so called line-by-line models for which the parameters are obtained from spectroscopic databases, such as high-resolution transmission molecular absorption database (HITRAN) \cite{HITRAN12}. The work herein utilizes the spectroscopic databases by obtaining the parameters for the major lines and we simplify their impact to be modelled by simple polynomials that only depend on the water vapor content in the air. These are then applied to the Beer-Lambert's law to obtain distance dependent absorption loss. The free space propagation is modelled by the square-law free space path loss (FSPL). Thus, the produced model is a simple and a relatively compact way to estimate the total free space loss on the above 100 GHz frequencies. The main use case of the produced model is to be able to omit the complicated spectroscopic databases that take efforts to implement and use flexibly. This is especially the case with the common wireless communications problems where detailed information on the source of the loss is not required, but just a way to model it easily.

Starting from the 100 GHz frequency, the six modelled molecular absorption lines are at about 119 GHz, 183 GHz, 325 GHz, 380 GHz, 439 GHz, and 448 GHz. This adds two lines at 119 GHz and 183 GHz to our previous model (\cite{Kokkoniemi2019}) in order to address the D band (110--170 GHz) propagation. The D band is among the next interesting range of frequencies for B5G systems.
The water vapor is the main cause of the absorption losses in the above 100 GHz frequencies such that all but one of the above lines are caused my it. The line at 119 GHz is caused by oxygen and it is comparably weak. Although weak, it has been included in the model, since it is part of the D band and as it causes a small attenuation in long distance links.

There are some existing works on parametric absorption loss models. International Telecommunication Union Radio Communication Sector (ITU-R) has provided a model to calculate gaseous attenuation up to 1000 GHz in ITU-R P.676-8 \cite{ITU676}. This model is line-by-line based and the its results are therefore matched with those by full spectroscopic databases. There is a difference to the proposed work; ITU-R uses a modified full Lorentz line shape that is not in general recommended for the millimeter frequencies \cite{Paine2012} due to heavy tailed frequency domain absorption distribution. A better choice is a  model that takes into account lower wing absorption by using line shape such as van Vleck--Weisskopf or van Vleck--Huber \cite{Paine2012}. Furthermore, the full model by ITU-R still requires large numbers of tabulated parameters (553) that render its utilization similarly slow as the full databases. In \cite{ITU676}, a polynomial based approximation is also given. It is valid up to 350 GHz, but it is somewhat usable up to about 450 GHz. Newer version of this model, ITU-R 676-11, also exists but that version does not have a polynomial model.  We use the older version in this paper as we present a similar (but more simpler) polynomial model.

Compared to the proposed model, the ones presented in \cite{ITU676} have several weaknesses. The ITU-R models \cite{ITU676} includes lines even up to 1780 GHz, but it is only specified to be valid for frequencies up to 350 GHz. The simplified model in the newer version is also limited to 350 GHz. The model also includes nine polynomials. If some of these terms are removed, they may also affect frequencies in different bands due to additive nature of the absorption lines. For example, the term involving 1780 GHz has to be kept or the attenuation levels between the peaks absorption frequencies at lower frequencies is incorrect. However, the ITU-R models are fairly accurate below 450 GHz. Because of the Full Lorentz line shape model, they overestimate the absorption line wing absorption. As detailed above, we will give a model with the extended frequency range and more accurate estimate for the absorption loss in simple form. This model can also be reduced to a simpler one (due to utilization of a fit parameter) for a desired sub-band within the full range of the model (100--450 GHz).

We have given a simplified molecular absorption loss model in the past in \cite{Kokkoniemi2018}. It was intended for the 275--400 GHz band. We also gave an extended version of that in \cite{Kokkoniemi2019} for frequencies from 200-450 GHz. This paper is an extended version of \cite{Kokkoniemi2019} with new lines focusing on the D band. As in the previous versions, the proposed model is based on the Lorentz absorption line shape. As mentioned in the context of ITU-R models, the Lorentz line shape overestimates the wing absorption. The correct line shape obtained using the van Vleck--Huber line shape is achieved by using a fit function. This fit function also ensures that the model gives correct output even if all the lines are not utilized. The proposed model is shown to be very accurate by numerical results in Section \ref{section:4}, where it is benchmarked against the actual response as well as the ITU-R models.

The rest of this paper is organized as follows: Section \ref{section:3} derives the proposed absorption loss model, Section \ref{section:4} gives some numerical examples, and Section \ref{section:5} concludes the paper. 

%% file: EUCNC19simpl.tex
\subsection{Molecular absorption loss}

The main goal of this paper is to describe the molecular absorption loss. The free space loss is given by a standard Friis equation. The molecular absorption is described by the Beer-Lambert law. It gives the transmittance, i.e., the fraction of energy that propagates through the medium at link distance $d$. This exponential power law depends on the link distance and absorption coefficient by \cite{Jornet2011,Paine2012}
\begin{equation}
\label{equation:tau_tot}
\tau(f,d) = \frac{P_r(f)}{P_t(f)} = e^{-\Sigma_{j}\kappa_\text{a}^j(f)d},
\end{equation}
where $\tau(f,d)$ is the transmittance, $f$ is the frequency, $d$ is the distance from transmitter (Tx) to receiver (Rx) (in meters in the models presented in this paper), $P_t(f)$ and $P_r(f)$ are Tx and Rx power, respectively, and $\kappa_\text{a}^j(f)$ is the absorption coefficient of the $j$th absorbing species at frequency $f$. The absorption coefficient is usually calculated with databases of spectroscopic parameters, such as the HITRAN database \cite{HITRAN12}. The detailed calculation of the absorption coefficient can be found, e.g., in \cite{Kokkoniemi2018}.

\subsection{Simplified absorption loss model}

The polynomial absorption loss model is obtained by searching the strongest absorption lines on the band of interest and extracting the parameters for those. The temperature and pressure dependent coefficients are fixed. Since the absorption on the frequencies above 100 GHz are mainly caused by the water vapor in the air, the volume mixing ratio of water vapor is left floating. The parametric model is characterized by the absorption coefficients $y_i$ at absorption lines $i$. The above Beer-Lambert model becomes
\begin{equation}
\label{eq:PLabs}
\text{PL}_\text{abs}(f,\mu) = e^{d\left(\sum\limits_iy_i(f,\mu) + g(f,\mu)\right)},
\end{equation}
where $f$ is the desired frequency grid, $y_i$ is an absorption coefficient for the $i$th absorption line, $g(f,\mu)$ is a polynomial to fit the  expression to the actual response (see below for more details), and $\mu$ is the volume mixing ratio of water vapor. It is determined by the relative humidity $\phi$ at temperature $T$ and pressure $p$ by
\begin{equation}
\mu = \frac{\phi}{100} \frac{p_w^\ast(T,p)}{p},
\end{equation}
where $\phi p_w^\ast(T,p)/100$ is the partial pressure of water vapor and $p_w^\ast$ is the water vapor partial pressure. This can be obtained, e.g., from the Buck equation \cite{Alduchov1995}
\begin{equation}
p_w^\ast = 6.1121(1.0007 + 3.46\times 10^{-6}p)\exp\left(\frac{17.502T}{240.97+T}\right),
\end{equation}
where the pressure $p$ is given in hectopascals and $T$ is given in degrees centigrade.

The six polynomials for the six major absorption lines at the 100--450 GHz band are the following\footnote{Notice that in the original EuCNC'19 paper \cite{Kokkoniemi2019} for which this paper is an extension to, there was an error that is rectified here. The terms $(f/100c-p_x)^2$ were not squared. This causes the model therein to give an incorrect output. However, at so notable level that it should be obvious if one tries to implement the model and compares to our results. The results in \cite{Kokkoniemi2019} were made with correct expressions.} :
\begin{equation}
y_1(f,\mu) = \frac{A(\mu)}{B(\mu) + \left(\frac{f}{100c}-p_1\right)^2},
\end{equation}
\begin{equation}
y_2(f,\mu) = \frac{C(\mu)}{D(\mu) + \left(\frac{f}{100c}-p_2\right)^2},
\end{equation}
\begin{equation}
y_3(f,\mu) = \frac{E(\mu)}{F(\mu) + \left(\frac{f}{100c}-p_3\right)^2},
\end{equation}
\begin{equation}
y_4(f,\mu) = \frac{G(\mu)}{H(\mu) + \left(\frac{f}{100c}-p_4\right)^2},
\end{equation}
\begin{equation}
y_5(f,\mu) = \frac{I(\mu)}{J(\mu) + \left(\frac{f}{100c}-p_5\right)^2},
\end{equation}
\begin{equation}
y_6(f,\mu) = \frac{K(\mu)}{L(\mu) + \left(\frac{f}{100c}-p_6\right)^2},
\end{equation}
\begin{equation}
g(f,\mu) = \frac{\mu}{0.0157}(2\times10^{-4}+af^b),
\end{equation}
where the frequency $f$ is given in Hertz, and
\begin{equation}
\nonumber
A(\mu) = 5.159\times 10^{-5}(1-\mu)(-6.65\times 10^{-5}(1-\mu) + 0.0159),
\end{equation}
\begin{equation}
\nonumber
B(\mu) = (-2.09\times 10^{-4}(1-\mu) + 0.05)^2,
\end{equation}
\begin{equation}
\nonumber
C(\mu) = 0.1925\mu(0.1350\mu + 0.0318),
\end{equation}
\begin{equation}
\nonumber
D(\mu) = (0.4241\mu + 0.0998)^2,
\end{equation}
\begin{equation}
\nonumber
E(\mu) = 0.2251\mu(0.1314\mu + 0.0297),
\end{equation}
\begin{equation}
\nonumber
F(\mu) = (0.4127\mu + 0.0932)^2,
\end{equation}
\begin{equation}
\nonumber
G(\mu) = 2.053\mu(0.1717\mu + 0.0306),
\end{equation}
\begin{equation}
\nonumber
H(\mu) = (0.5394\mu + 0.0961)^2,
\end{equation}
\begin{equation}
\nonumber
I(\mu) = 0.177\mu(0.0832\mu + 0.0213),
\end{equation}
\begin{equation}
\nonumber
J(\mu) = (0.2615\mu + 0.0668)^2,
\end{equation}
\begin{equation}
\nonumber
K(\mu) = 2.146\mu(0.1206\mu + 0.0277),
\end{equation}
\begin{equation}
\nonumber
L(\mu) = (0.3789\mu + 0.0871)^2,
\end{equation}
and with $p_1 = 3.96$ cm$^{-1}$, $p_2 = 6.11$ cm$^{-1}$, $p_3 = 10.84$ cm$^{-1}$, $p_4 = 12.68$ cm$^{-1}$, $p_5 = 14.65$ cm$^{-1}$, $p_6 = 14.94$ cm$^{-1}$, $a = 0.915\times 10^{-112}$, $b = 9.42$. The lines $y_1$, $y_2$, $y_3$, $y_4$, $y_5$, and $y_6$ correspond to strong absorption lines at center frequencies 119 GHz, 183 GHz, 325 GHz, 380 GHz, 439 GHz, and 448 GHz, respectively. This is also visible in the line expressions as the parameters $p_1$ to $p_6$ give the line center frequencies in wavenumbers.

The above absorption lines were estimated based on the simple Lorentz line shape. The reason is the simpler form as compared to more accurate, but at the same time more complex line shapes, such as the van Vleck--Huber line shape. This produces an error as the Lorentz line shape over estimates the absorption line wing absorption. Therefore, the fit polynomial $g(f,\mu)$ is introduced. This fit polynomial also takes care of the wing absorption in the case the model is only utilized partially. That is, if one only utilizes some of the lines to model a subband within the full 100--450 GHz band, the same fit polynomial as in full model should always be included. It was obtained by curve fitting to the difference between the exact response and the response of the above $y_i$ lines. It would be possible to calculate the exact difference theoretically, but would only apply to the in-band absorption lines and this would not consider the out-of-band wing absorption, mainly from lines above 450 GHz. The total absorption loss with the above model is shown to produce very accurate estimate of the loss in the numerical results.

\begin{figure}[t!]
    \centering
    \includegraphics[width=3.1in]{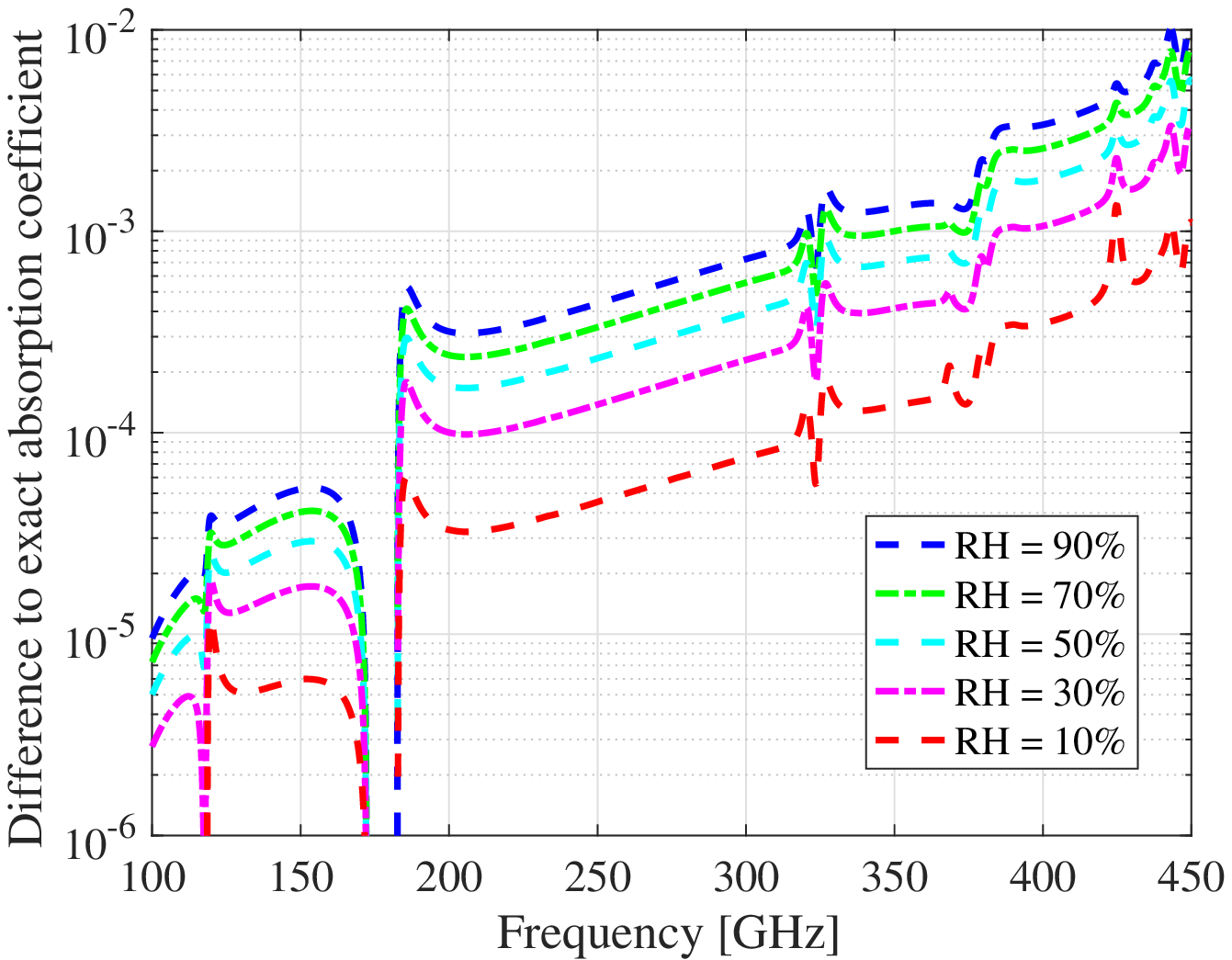}
    \caption{An error of the absorption coefficient of the proposed model to the exact one as a function of the frequency for different humidity levels before adding the fit polynomial $g(f,\mu)$.}
    \label{fig:Fitting}
\end{figure}

The water vapor volume mixing ratio is taken into account in the fit polynomial $g(f,\mu)$. Whereas it is highly accurate, this estimate will cause some error that is dependent on the water vapor level. Fig. \ref{fig:Fitting} shows the error of the absorption coefficient to the exact one based on the above absorption loss model and before applying the fit polynomial. This error was calculated at 25 degrees centigrade and in various humidity levels $\mu$ = [0.0031 0.0094 0.0157 0.0220 0.0282] that correspond to relative humidities $\phi$ = [10\% 30\% 50\% 70\% 90\%], respectively. Notice the exponential y-axis, the error is not very large. However, the error increases as a function of frequency. This is due to the increasing and accumulating wing absorption from the higher frequency lines. This is the error the fit polynomial $g(f,\mu)$ rectifies by adjusting the absorption lines shapes. The 0.0157 in $g(f,\mu)$ comes from the design atmospheric conditions of 25 degrees Celsius and 50\% relative humidity. It should be noticed that in the total absorption line, the error is the smallest for low humidity just due to the fact that there is less water in the air, and thus, the overall difference between the exact and estimated absorption coefficient is low to begin with.

\subsection{FSPL and the total loss}

The FSPL is given by the Friis transmission equation:
\begin{equation}
\label{eq:FSPL}
\text{PL}_\text{FSPL}(d,f) = \frac{(4\pi d f)^2}{c^2},
\end{equation}
where $c$ is the speed of light. We focus herein only on the free space propagation and thus the total loss is given by composition of the FSPL and the molecular absorption loss as
\begin{equation}
\label{eq:totloss}
\text{PL}(d,f) = \frac{(4\pi d f)^2\exp(\kappa_\text{a}(f,\mu)d)}{c^2}G_\text{Rx}G_\text{Tx},
\end{equation}
where $G_\text{Rx}$ and $G_\text{Tx}$ are the antenna gains. When using the polynomial models above, the absorption coefficient $\kappa_\text{a}(f,\mu)$ is
\begin{equation}
\label{eq:kappares}
\kappa_\text{a}(f,\mu) = \sum\limits_iy_i(f,\mu) + g(f,\mu),
\end{equation}
where the $y_i(f,\mu)$ are the above polynomial absorption lines (and as also shown in (\ref{eq:PLabs})), or subset of those depending on the modelled frequency band within the frequency range from 100--450 GHz. For instance, D band propagation would only need lines  $y_1(f,\mu)$ and $y_2(f,\mu)$. Another popular band for high frequency communications is the 275--325 GHz band. Then only the line $y_3(f,\mu)$ would be enough. The fit polynomial $g(f,\mu)$ is always required and because of it we can use very low complexity models for the subbands, further pronouncing the complexity benefits as compared to the ITU-R polynomial model. It will be shown in the numerical results that these subsets give very accurate estimate of the loss also in partial bands without a need to implement all the lines in the model.

%% file: EUCNC19results.tex
In this section, we first present some performance analysis of the proposed absorption loss model. This is done by analysing the error produced by the model to the exact model, as well as to the ITU-R parametric and full models. Secondly we give some numerical examples of the model and path loss outputs.

\subsection{Performance analysis}

\begin{figure}[t!]
    \centering
    \includegraphics[width=3.1in]{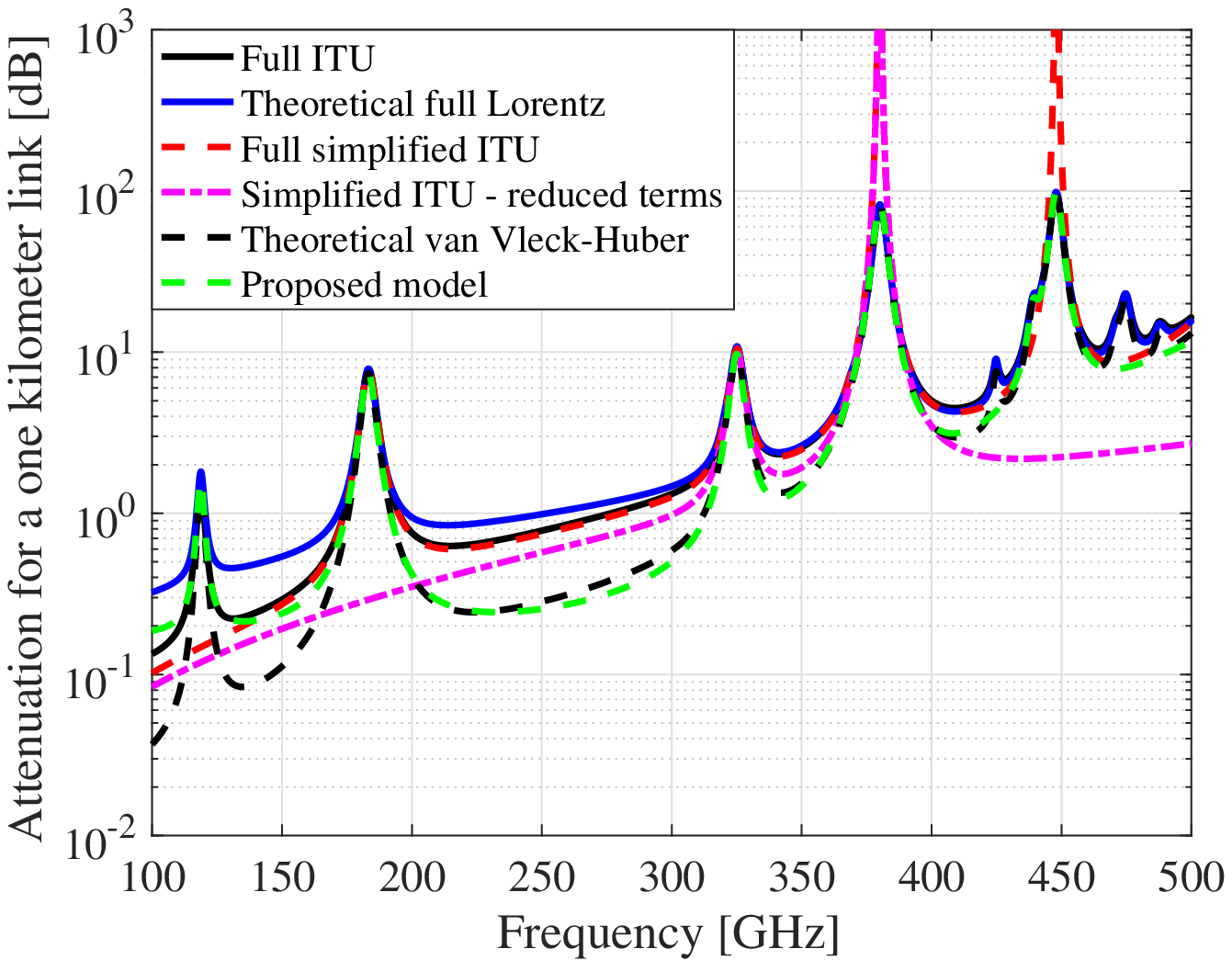}
    \caption{Molecular absorption loss at 1 km distance at 25 degrees centigrade and 10\% relative humidity ($\mu = 0.0031$).}
    \label{fig:Loss10RH}
\end{figure}

\begin{figure}[t!]
    \centering
    \includegraphics[width=3.1in]{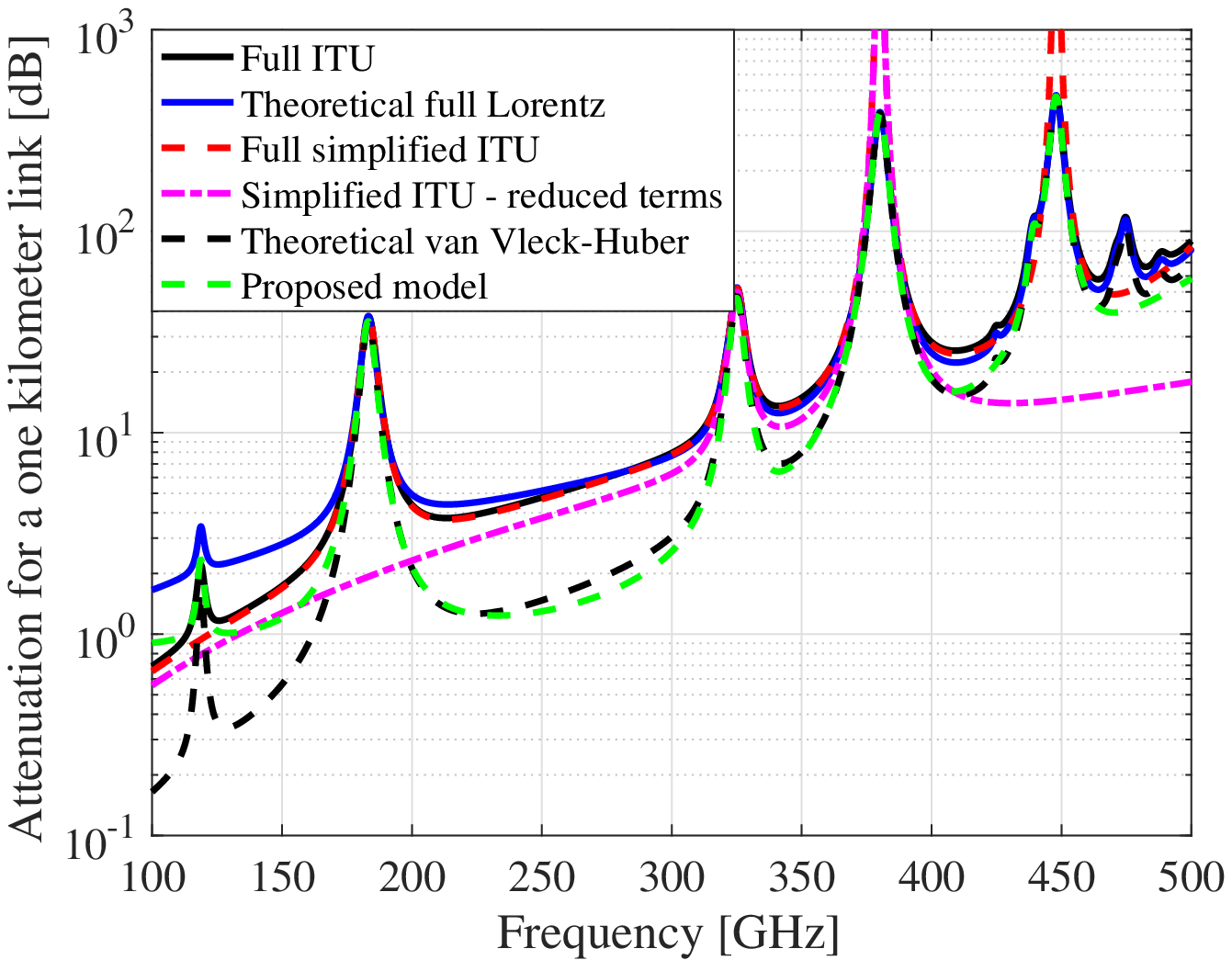}
    \caption{Molecular absorption loss at 1 km distance at 25 degrees centigrade and 50\% relative humidity ($\mu = 0.0157$).}
    \label{fig:Loss50RH}
\end{figure}

\begin{figure}[t!]
    \centering
    \includegraphics[width=3.1in]{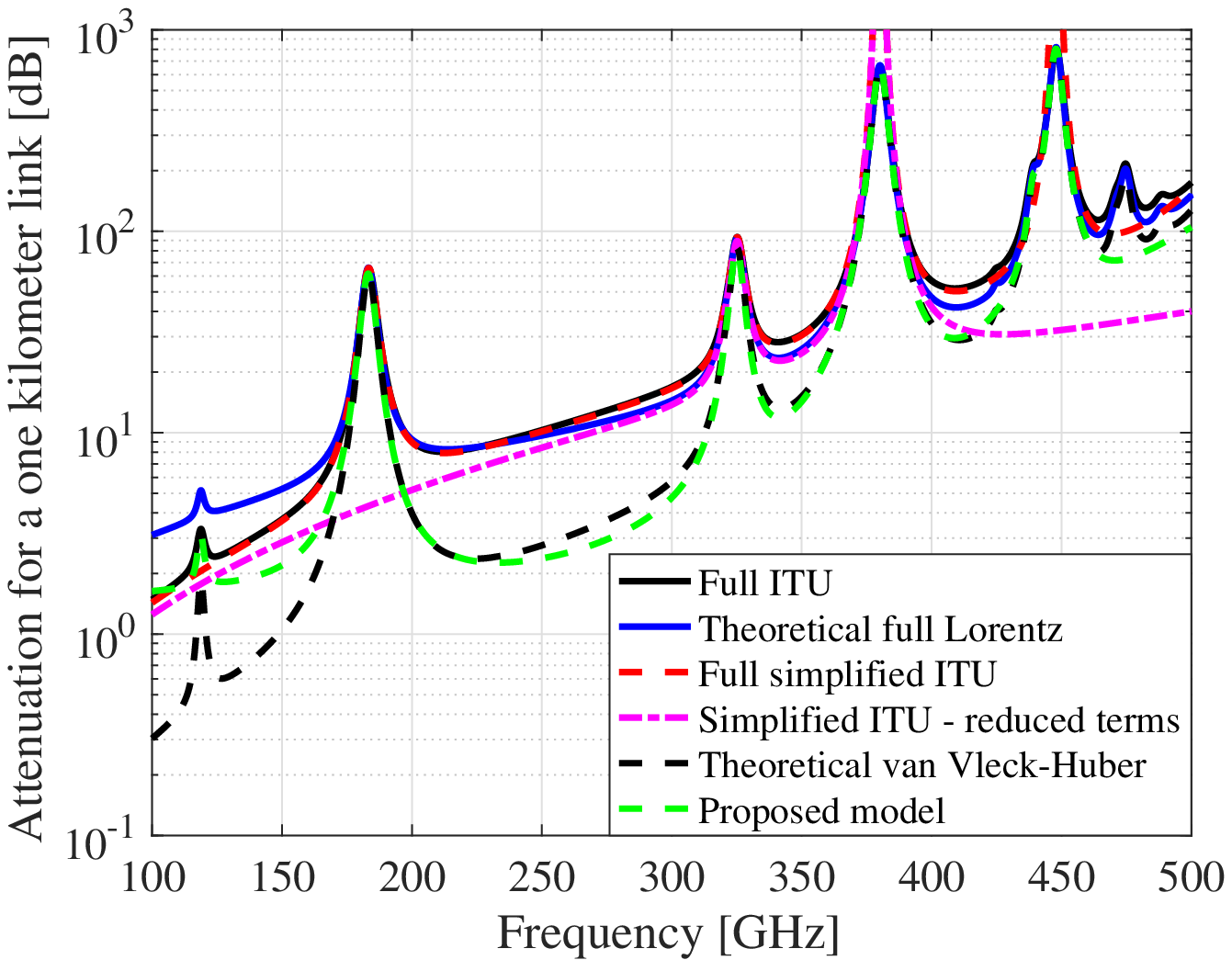}
    \caption{Molecular absorption loss at 1 km distance at 25 degrees centigrade and 90\% relative humidity ($\mu = 0.0282$).}
    \label{fig:Loss90RH}
\end{figure}

We compare the path loss values of the proposed molecular absorption loss model versus the ITU-R models in Figs.\ \ref{fig:Loss10RH} to \ref{fig:Loss90RH} for the relative humidity levels from 10\% to 90\%, respectively, at 25 degree centigrade for a one-kilometer link. A high link distance was used to emphasize the differences between the models. This is because the impact of the molecular absorption loss decreases for short distances due to exponential power law.

As it was predicted above, the Lorentz line shape (along with the full Lorentz line shape) overestimates the wing absorption. This is not a major issue at higher parts of the THz band due to more lines and line mixing. However, at the lower frequencies this is a problem because the Lorentz line shape does not attenuate the response fast enough towards the zero frequency. As a consequence, the ITU-R models give higher path loss figures in general for below 500 GHz frequencies. The difference to the actual response vary from few dBs to tens of dBs depending on the link distance and humidity level. Notice that the simplified reduced version of the ITU-R model does not include all the lines rather than being incorrect.

\begin{figure}[t!]
    \centering
    \includegraphics[width=3.1in]{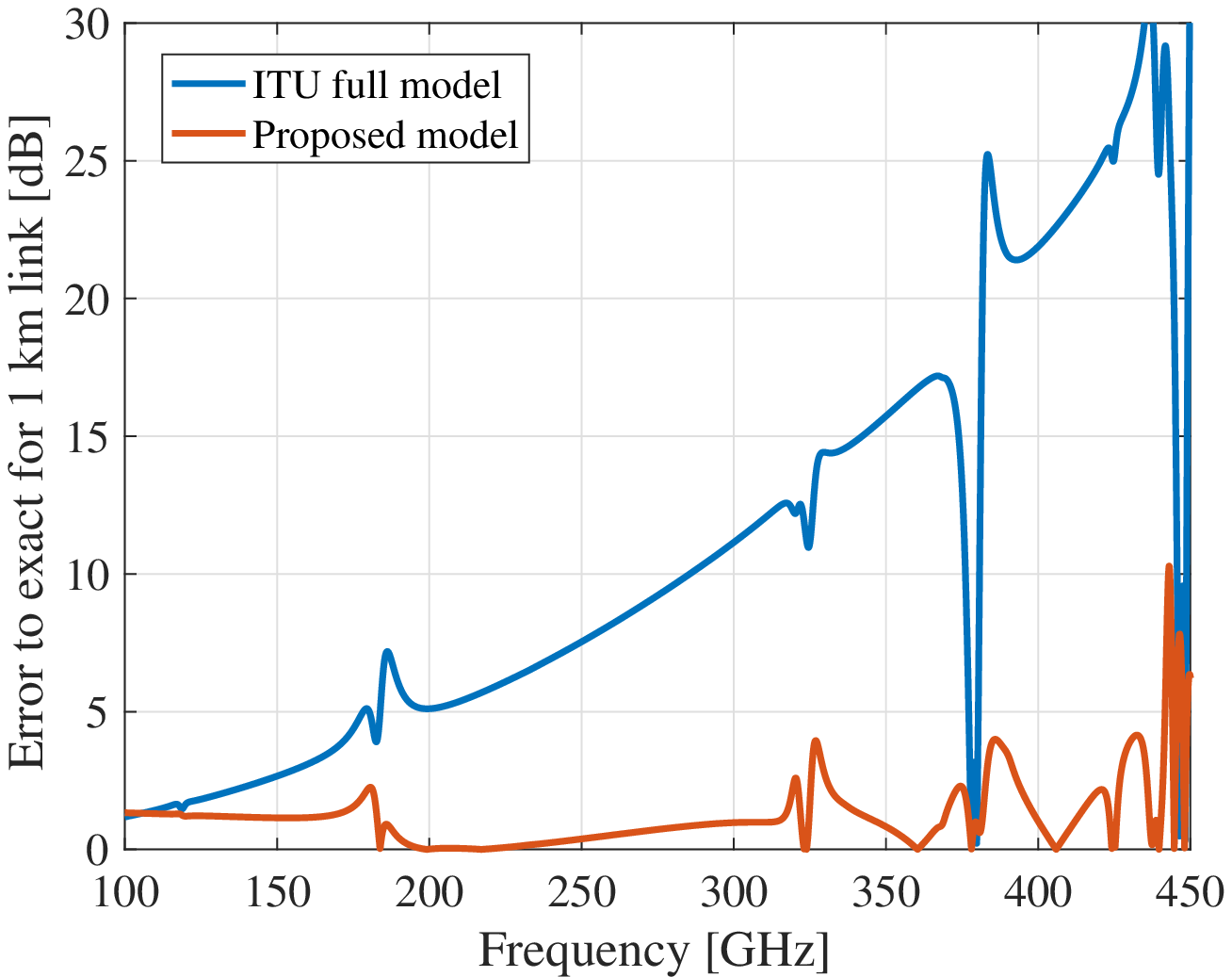}
    \caption{Absolute errors given by the ITU-R full model and the proposed model to the exact theory for a one kilometer link.}
    \label{fig:Error}
\end{figure}

There are a couple of further observations to be made. The ITU-R models are based on the full Lorentz model, but the database specific one does overestimate the response even more. This is due to reason that the ITU-R model is a modified version of the full Lorentz model that increases its accuracy. Second observation is that the proposed model is rather accurate, but not perfect. In Figs.\ \ref{fig:Loss10RH} to \ref{fig:Loss90RH} the difference seems to be largest below 200 GHz. However, the large part of the large apparent difference comes from the logarithmic y-axis. Fig.\ \ref{fig:Error} gives the true worst case error herein. This figure shows the error of the path loss for one kilometer link at 25 degree Celsius and at 90\% relative humidity. It can be seen that the error is very good across the board, but the lower frequencies do give comparably slightly larger error due to in general lower absorption loss. However, the figures herein are for one kilometer link and the error will decrease with decreasing distance due to exponential behavior of the absorption loss. Thus, the resultant error of roughly $\pm$2 dB is very good for such extremely high link distances considering the high frequencies and their general applicability to low range communications. Furthermore, the error also decreases in less humid environment and this is in general true for ITU-R models as well. For instance at 10\% relative humidity at 25 degree Celsius, the differences are rather modest. Regardless if this, in more humid and therefore more realistic conditions there is a notable difference between the models, especially above the 200 GHz frequencies. 

As a last note on the error performance, all the models herein are rather accurate and it is an application specific issue how accurately the absorption loss needs to be calculated. If the link distance is high or the communications band is in the vicinity of the absorption line, the importance of the correct loss is high. However, on low distance links and in the middle of the low loss regions of the spectrum the absorption loss is modest and large error is not made if the absorption loss is omitted altogether.

\begin{figure}[t!]
    \centering
    \includegraphics[width=3.1in]{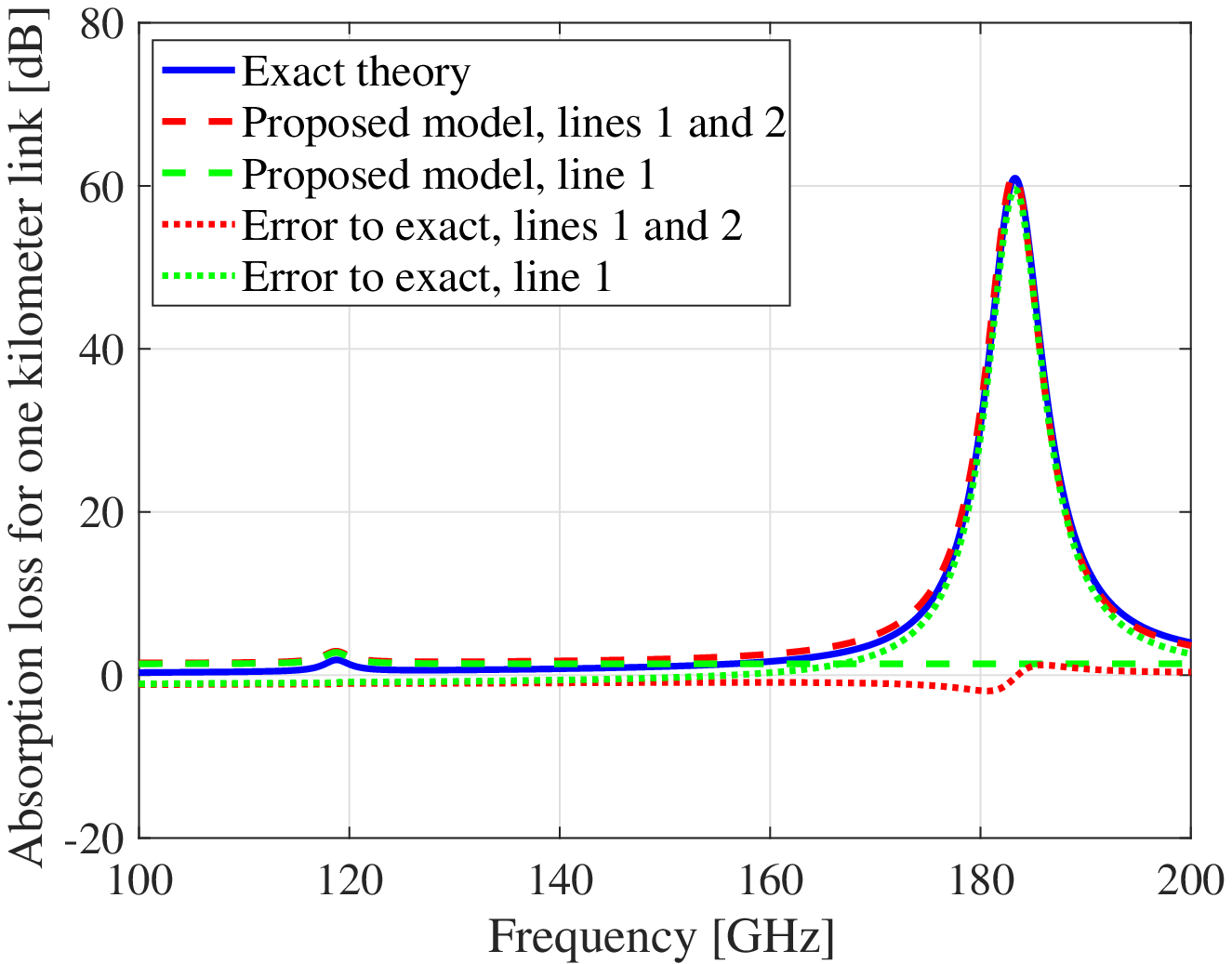}
    \caption{Reduced versions of the proposed model giving absorption losses up to about 160 GHz (1 term) and 200 GHz (2 terms).}
    \label{fig:LossReduced}
\end{figure}

\begin{figure}[t!]
    \centering
    \includegraphics[width=3.1in]{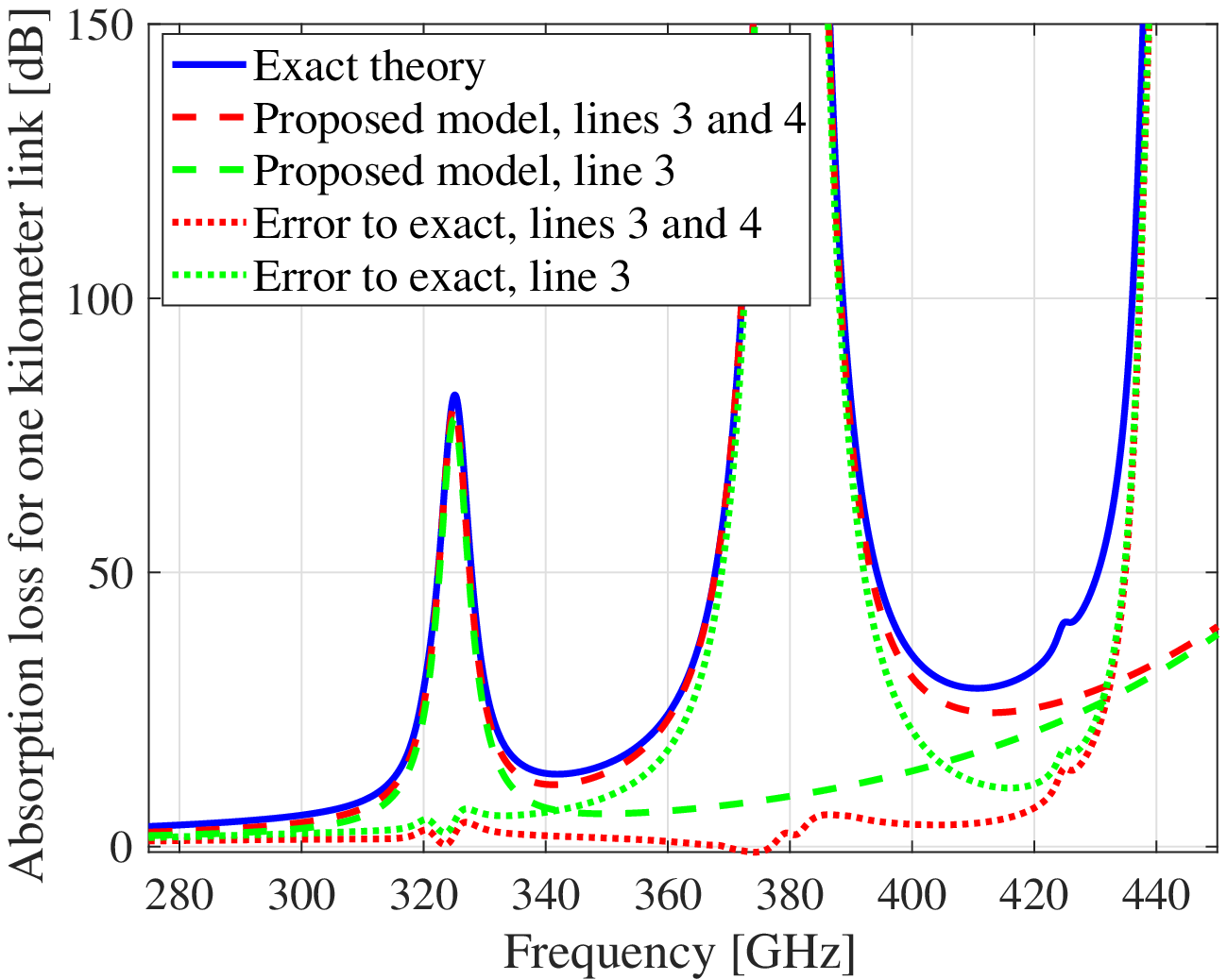}
    \caption{Reduced versions of the proposed model giving absorption losses up to about 330 GHz (3rd line only) and 390 GHz (lines 3 and 4 only).}
    \label{fig:LossReduced2}
\end{figure}

Figs.\ \ref{fig:LossReduced} and \ref{fig:LossReduced2} compare the performance of the proposed model with reduced terms against the exact theory. Fig.\ \ref{fig:LossReduced} shows performance of the proposed model with the first two lines separately and jointly at about 119 GHz and 183 GHz (shown as lines 1 and 2 in the figure). In the other words, one should utilize the absorption coefficient as $\kappa_\text{a}(f,\mu) = y_1(f,\mu) + g(f,\mu)$ or $\kappa_\text{a}(f,\mu) = y_1(f,\mu) + y_2(f,\mu) + g(f,\mu)$ for lines 1 and 1 and 2 jointly, respectively. This reduction would correspond roughly to channel models at the D band. It can be seen that the proposed model with reduced terms performs very well on estimating the absorption loss. The same occurs in the case of Fig.\ \ref{fig:LossReduced2} that shows the performance of the next two lines (lines 3 and 4) corresponding to frequencies 325 GHz and 380 GHz. These two line alone give a very good estimate of the loss up to about 330 GHz and 390 GHz for the line 3 and joint lines 3 and 4, respectively. These correspond to utilizing an absorption coefficient as $\kappa_\text{a}(f,\mu) = y_3(f,\mu) + g(f,\mu)$ and $\kappa_\text{a}(f,\mu) = y_3(f,\mu) + y_4(f,\mu) + g(f,\mu)$ for lines 3 and 3 and 4 jointly. As such, the line 3 would be mostly enough for the popular transition frequencies between the mmWave and THz bands. Namely 275--325 GHz. However, with these two lines, the model remains accurate from about 200 GHz up to the above mentioned 390 GHz. Therefore, the proposed model is flexible and easily reducible for multiple frequency band within the full range from 100 to 450 GHz.

\subsection{Path loss examples}

\begin{figure}[t!]
    \centering
    \includegraphics[width=3.2in]{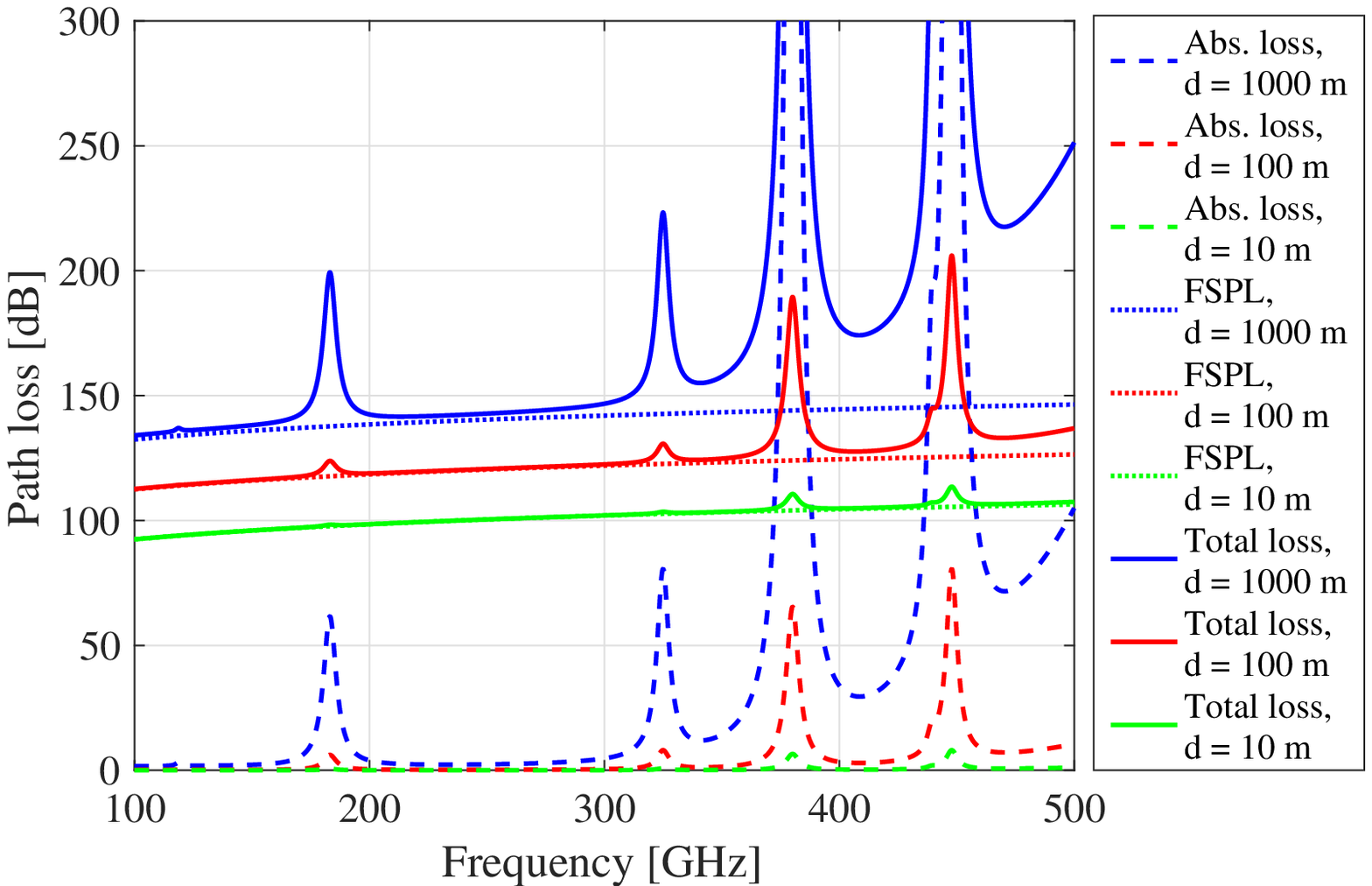}
    \caption{Molecular absorption loss versus FSPL as a function of frequency for few distances.}
    \label{fig:TotalLoss}
\end{figure}

Fig.\ \ref{fig:TotalLoss} shows the total attenuation values as a function frequency for the FSPL, molecular absorption loss and the total loss assuming unit antenna gains. As mentioned above, we can see that the low distances impose minimal absorption impact and the dominant loss comes from the FSPL. Increasing the distance causes the absorption to become a dominant loss. This is due to square power law versus exponential power law and how those behave as a function distance. Thus, in the low range applications the error is not large of the absorption is omitted. On the other hand, some backhaul applications require absorption to be taken into account for correct link budget calculations. The absorption also has impact on the 3-dB bandwidths in between the absorption lines and in close vicinity of those. We discussed those in detail and gave a simple model for the estimated band in the first paper of the simplified molecular absorption loss model in \cite{Kokkoniemi2018}. The need to model the absorption loss depends on the application, but where it is needed, simple models, such as one presented herein, give easy way to achieve accurate estimate for it.

\begin{figure}[t!]
    \centering
    \includegraphics[width=3.2in]{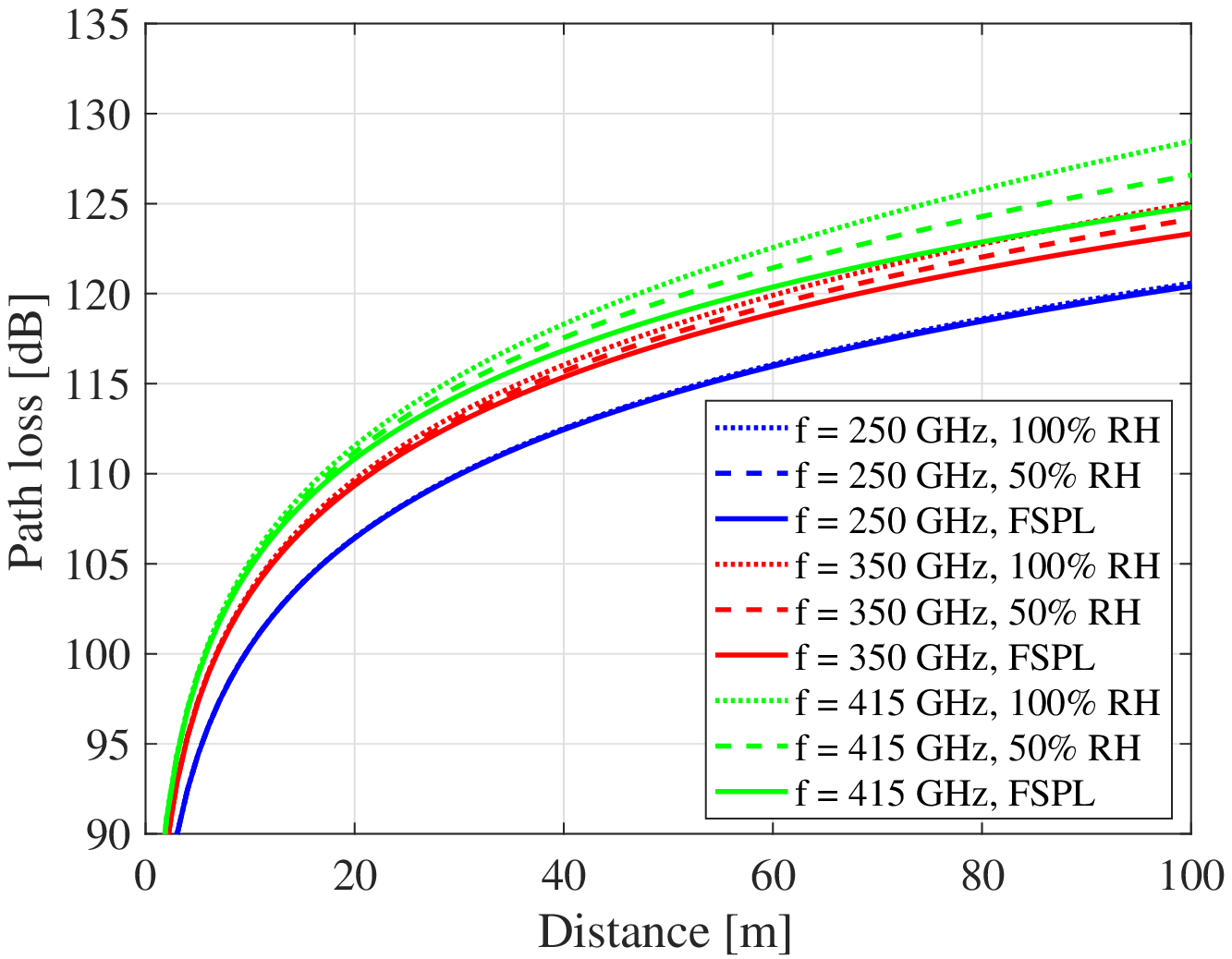}
    \caption{Path loss as a function of distance for 5 GHz bandwidth for three different center frequencies. Losses are given for 30 degree ambient temperature and for 50\% and 100\% relative humidities, and for the pure FSPL.}
    \label{fig:PL5}
\end{figure}

\begin{figure}[t!]
    \centering
    \includegraphics[width=3.2in]{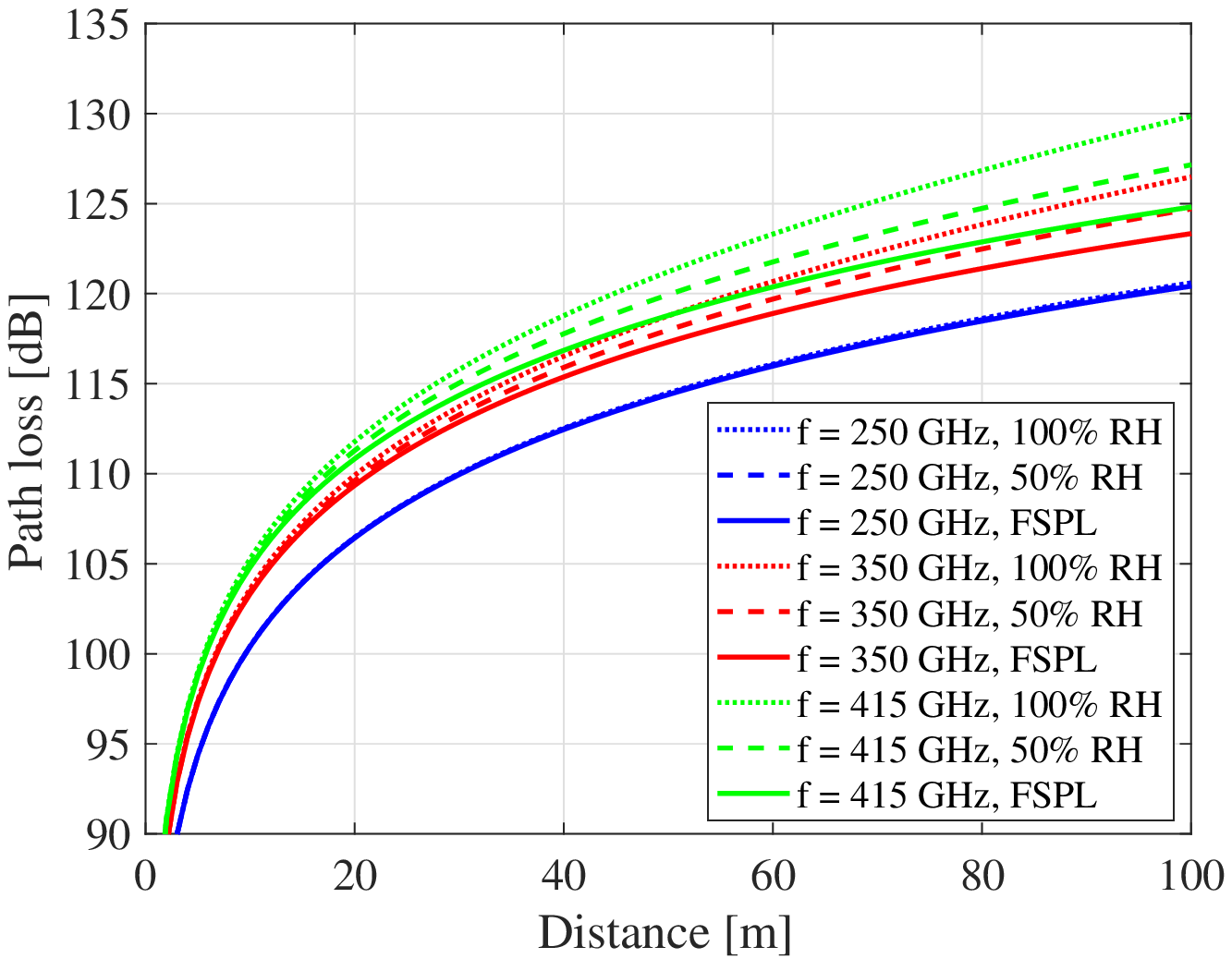}
    \caption{Path loss as a function of distance for 40 GHz bandwidth for three different center frequencies. Losses are given for 30 degree ambient temperature and for 50\% and 100\% relative humidities, and for the pure FSPL.}
    \label{fig:PL40}
\end{figure}

Figs.\ \ref{fig:PL5} and \ref{fig:PL40} show the average path loss for 5 GHz and 40 GHz signal bandwidths, respectively, at three center frequencies: 250 GHz, 350 GHz, and 415 GHz. These frequencies correspond to some of the low loss regions between the absorption lines, i.e., the desirable operational band for real applications as it would be preferable to avoid by design the high loss regions. These figures give the losses as a function of distance and with an without the molecular absorption loss. The FSPL is identical in the both figures. The above mentioned bandwidth shrinking as a function of distance in close vicinity of the absorption lines has an impact on the total path loss. This is expected and visible in two ways: 1) by comparing the 5 GHz path losses to the corresponding 40 GHz ones, 2) the increased humidity increases the absorption loss and causes larger loss. The latter though is partially due to increased wing absorption that causes increased loss also between the lines. Overall picture is that the path loss increases in the THz regime due to the molecular absorption loss.

\begin{figure}[t!]
    \centering
    \includegraphics[width=3.2in]{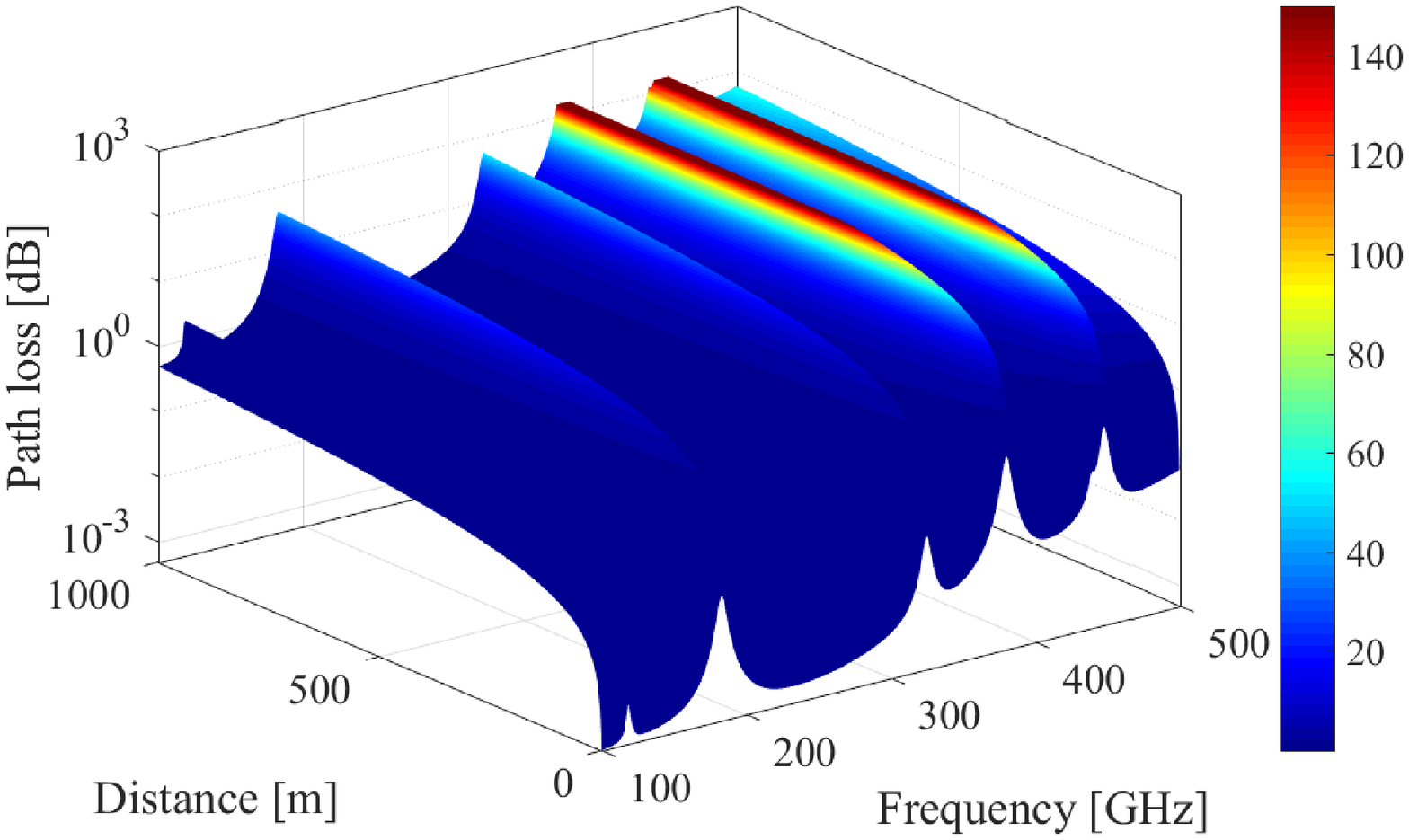}
    \caption{3D surface plot for the total loss as a function of frequency and distance according to the proposed model in 23 degree ambient temperature and 50\% relative humidity.}
    \label{fig:3DPL50}
\end{figure}

Finally, Fig.\ \ref{fig:3DPL50} gives the absorption loss in the considered frequency range and as a function of distance. Notice that the loss has been suppressed to have a maximum loss of 150 dB due to quickly and exponentially aggregating attenuation at the absorption line centers. This figure also shows the picture of the above 100 GHz communications. The expected losses increase fast with distance, but there are still many potential frequency regimes where even high range communications is possible. The FSPL is always there and causes significant losses on the signals. This forces to always use high gain antennas at high frequency communications. In general though, the above 100 GHz frequencies show and have shown great potential to serve as one of the opportunistic platforms for the future B5G systems.

\subsection{Discussion}

As a summary and discussion from above, the higher mmWave and low THz frequencies are among the most potential frequencies to utilize ultrahigh rate communications in the future wireless systems. The proper modeling of the channel behavior therein is very important due to absorption loss and how it behaves in comparison to the FSPL. In low range communications it is not absolutely crucial to model the absorption due to dominating FSPL. Its importance increases with link distance, but also with frequency. In the other words, the link budget and the components in it are application dependent. The tools provided herein give and easy way to model the absorption loss and estimate its impact on the link budgets.

%% file: EUCNC19concl.tex
We derived a LOS channel model for 100--450 GHz frequency region. The main emphasis was on finding a molecular absorption loss with simple and easy to use model. This model was shown to be very accurate and predict the channel loss very well in the above frequency regime. The derived model can also be reduced to even simpler form, if one is interested in some particular frequency band within the whole range. Considering the upcoming B5G systems, such bands include the D band (110 GHz to 170 GHz) and the transition band between the mmWave and the THz frequencies (275 GHz to 325 GHz). The molecular absorption loss is an important part of the link budget consideration for high frequency wireless communications. Therefore, the model presented here gives an easy way to estimate the total link loss in various environmental conditions and link distances, and therefore, in the B5G systems.


%% file: EUCNC19ack.tex
This work was supported by Horizon 2020, European Union's Framework Programme for Research and Innovation, under grant agreement no. 761794. This work was also supported in part by the Academy of Finland 6Genesis Flagship under grant no. 318927.